\documentstyle[aps,prl,multicol,epsf]{revtex}
\begin{document} 
\draft 
\title{Chiral Metal as a Heisenberg Ferromagnet} 
\author{H. Mathur$^{1,2}$} 
\address{$^{1}$Department of Physics,
Case Western Reserve University, Cleveland, Ohio
44106-7079} 
\address{$^{2}$Institute for Theoretical
Physics, University of California at Santa Barbara,
Santa Barbara, CA 93106-4030}

\date{\today} 
\maketitle


\begin{abstract}

The two-dimensional surface of an integer quantum hall multilayer
is mapped onto a Heisenberg spin-chain with ferromagnetic
coupling. Using 
this mapping it is shown non-perturbatively
that the surface states constitute a very
anisotropic metal in the infinite size limit.
For multilayers of finite size, two diffusive
mesoscopic regimes are identified and the
conductance fluctuations are calculated
perturbatively for both. The 
Heisenberg spin-chain representation is used to study
the directed wave problem and the exact result
is obtained that the mean-square deflection of a directed
wave grows as the square root of the propagation
distance.

\end{abstract} 
\pacs{PACS: 72.15.Rn; 75.10.Jm; 73.40.Hm; 42.25.Bs }

\begin{multicols}{2} 

Dirty electronic systems exhibit a variety of phases
many of which are not well-understood \cite{lee},
\cite{belitz}.  For
example electrons moving in two dimensions under the
influence of a magnetic field exhibit a sequence of
localization-delocalization transitions as the magnetic
field or some other system parameter is varied.
Understanding of these transitions, which underlie the
quantum hall effect, is based largely on numerical
simulation \cite{huckestein}.

The purpose of this Letter is to study the two
dimensional electronic states that live on the surface
of an integer quantum hall multilayer (sometimes called
the bulk quantum hall effect).   At the edge of each
quantum hall layer the electrons circulate in one sense
only and may be modelled as non-interacting chiral fermions
\cite{halperin}.  If the layers are coupled by
tunneling, the surface states comprise a
two-dimensional chiral electronic system (see fig 1).
The Bechgaard salts are
a natural realization \cite{chaikin} and it
is also possible to fabricate an appropriate semiconductor
heterostructure \cite{stormer}. 
It will be seen below that the chiral model used to
describe the surface states of the quantum hall
multilayer (eq 1 below) bears some similarity to a model
for the quantum hall effect introduced by Chalker and
Coddington \cite{chalker}.  However there is a crucial
difference which makes the chiral model tractable
whereas the Chalker-Coddington model has defied
solution.

The key question from the point of view of quantum
transport is whether the surface electronic wave functions are
localized or extended along the direction of the field
(the $z-$direction in fig 1).  This determines whether
the system is metallic or insulating since transport 
along the $z-$direction would be dominated by
the surface.  In either case, to fully
characterize the transport it is not sufficient to
study the disorder-averaged conductance: Localized
electronic systems generally possess a very broad
distribution of conductances. Although in contrast
metallic systems of ordinary (rather than chiral)
electrons do not have a broad conductance distribution,
the conductance fluctuations of finite-sized or
{\em mesoscopic} metallic grains have remarkable {\em
universal} properties \cite{stone} (for example, the
typical fluctuations are of order $ e^{2}/h
$$-$independent of the mean conductance or other
details of the sample).

Quantum transport in the chiral model has previously
been studied numerically \cite{daniell} and by mapping
onto field theories \cite{balents,kim,matthew}.  In
particular Balents {\em et al.}  \cite{matthew} have
mapped the system onto a one-dimensional supersymmetric
ferromagnetic spin-chain using the well-established
supersymmetric technique for performing disorder
averages \cite{efetov}.  Using this mapping they are
able to establish the important result that the 
two-dimensional chiral
model is metallic (in the sense that the conductance
scales ohmically) even for arbitrarily strong disorder.
This should be contrasted with conventional electronic
systems in which metallic scaling is associated with
large conductance and which are generically localized
in two dimensions \cite{lee}.  The surface states of
the bulk quantum hall system are thus revealed to be a
novel metallic phase with interesting
localization and mesoscopic fluctuation physics that
awaits exploration.

The chiral model (eq 1 below) that describes the
surface of a quantum hall multilayer is of interest
from another quite distinct point of view.  It should
apply whenever waves propagate in a medium that is
sufficiently anisotropic to warrant neglect of
backscattering in one direction.  The problem of waves
propagating in an anisotropic medium has been the focus
of much attention and is known as the directed wave
problem (see for example ref \cite{feng,saul} and
references therein).  Although directed waves are
described by the chiral model the question one asks in
this context is very different:  The electron is
assumed to be localized at a single point initially.
Thereafter it moves ballistically in the chiral
$x-$direction and the wave function spreads (presumably
diffusively) in the transverse $z-$direction.  The
interesting questions here concern the growth of the
wave packet width and the fluctuations in the position
of the centre of the wave-packet (denoted $ [ <n>^{2} ]_{{\rm imp}} $; 
the notation is explained below).  The broadening
of the width can be easily calculated and the answer
has been known since the 1970s
\cite{russian}.  $ [ <n>^{2} ]_{{\rm imp}} $ is more
difficult to calculate as the average of four green's
functions is now required; it has not been evaluated
within this model previously.

In this paper a new approach to disorder averaging is
introduced which is distinct from the conventional
replica or supersymmetry methods and is especially
adapted to this system.  Using this method it is
possible to map the chiral model (in the limit of
infinite size) onto a much simpler soluble
one-dimensional model:  an ordinary Heisenberg
ferromagnet.  The absence of localization for the
chiral system established by ref
\cite{daniell,balents,matthew} then emerges
non-perturbatively as a consequence of the well-known
quadratic dispersion of ferromagnetic magnons.  The
advantages of this mapping onto an ordinary ferromagnet
become evident when a more difficult calculation is
attempted.  For example $ [ <n>^{2} ]_{{\rm imp}} $ can be
expressed in terms of the matrix elements of two-magnon
states of the ferromagnetic representation.  Although
magnons interact, it is not difficult to obtain the
two-magnon eigenstates of a ferromagnet
\cite{bethe}.  Carrying
out such a calculation leads to the {\em exact} result
that $ [ <n>^{2} ]_{{\rm imp}} $ grows as $ x^{1/2} $. This
exact result agrees with the numerical simulations of
\cite{multifractal} but contradicts those of
\cite{feng2}; it also agrees with results obtained from
a simplified lattice model of directed wave propagation
introduced by Saul, Kardar and Read
\cite{saul,igor,me}.  In their model special
assumptions are made about the disorder which make it
possible to directly evolve the probability function
(the modulus square of the wave function) without
reference to the wave function itself.  The results of
this paper show that the same behaviour results for a
more generic disorder distribution.

Finally the properties of mesoscopic quantum hall
multilayers are studied.  Due to the anisotropy of the
surface states {\em two} diffusive mesoscopic regimes
can be identified.  Following the terminology of ref
\cite{matthew} in the one-dimensional diffusive regime
electrons are typically able to wind around the sample in the
chiral direction many times before diffusing across in
the $z-$direction.  The opposite limit of sufficiently
large circumference that winding paths are rare is
called the zero-dimensional regime.  The mesoscopic
regimes are difficult to analyse due to the winding
paths.  Spectral correlations were computed in ref
\cite{matthew} in the zero-dimensional limit; however
spectral correlations are difficult to probe
experimentally.  Here conductance fluctuations are
calculated in the two mesoscopic regimes using the
standard methods of impurity averaged perturbation
theory \cite{stone}.  Universal conductance
fluctuations of order $ e^{2}/h $ are found in the
one-dimensional limit coinciding with the result for
ordinary metals.  Interestingly, there is a crossover to much larger
fluctuations in the zero-dimensional diffusive limit.

The precise model used for the multilayer surface
is now described.
Each quantum hall layer has as many edge channels as
there are filled Landau levels.  For simplicity we
shall focus on the case of just one filled Landau
level.  The surface of the multilayer is then a two
dimensional chiral electronic system governed by the
Schr\"{o}dinger equation 
\begin{eqnarray} 
\left( - i v \frac{\partial }{ \partial x } + V_{n}(x) 
- E \right) G^{R}_{E} (n, x ; n', x') & & \nonumber \\
-t \{ G^{R}_{E} (n+1, x ; n', x') + G^{R}_{E} (n-1, x ; n',
x') \} & &  \nonumber \\
 = - i v \delta_{n, n'} \delta ( x - x' ). & &
\end{eqnarray} Here $ G^{R}_{E} $ is the retarded
Green's function at frequency $ E $. $ t $ produces interlayer
hopping and $ V_{n} $ is the disorder potential. Units
are chosen so that the edge velocity $ v = 1 $,
the interlayer separation $ a = 1 $ and $ \hbar = 1 $.
The anisotropy and chiral character of the model
are reflected by the fact that the equation is
first order in the chiral $x-$ direction whereas
it is second order in the transverse direction.

A mapping onto a one-dimensional problem is obtained
by noting that in the limit of infinite size (but in
that limit only) the circumference 
$ C \rightarrow \infty $ and the retarded
Green's function obeys the chiral boundary condition
$ G^{R}_{E}(n,x;n',x') = 0 $ for $ x < x' $.
Due to this boundary condition, it is possible to interpret
eq (1) as the {\em time-dependent} Schr\"{o}dinger
equation for a one-dimensional tight-binding model
with the chiral co-ordinate $ x $ identified as time
and $ G^{R}_{E}(n,x,;n',x') $ identified as the 
time-domain retarded Green's function. Note that
the on-site energies of the tight-binding model
fluctuate in time.
It is convenient to make a gauge
transformation $ G^{R} 
\rightarrow G^{R} \exp[ i \gamma_{n}(x) - i \gamma_{n'}(x') ] $
where $ \partial \gamma_{n} (x) / \partial x = V_{n} ( x ) - E $.
In this gauge eq (1) becomes
\begin{eqnarray}
- i \frac{ \partial }{ \partial x } G^{R}_{E} ( n, x; n', x' )
- t_{n}( x ) G^{R}_{E} (n + 1, x; n', x')  & & \nonumber \\
- t^{*}_{n-1}(x)
G^{R}_{E}(n - 1, x; n', x') = - i \delta_{n,n'} \delta( x - x' ). & &
\end{eqnarray}
Here $ t_{n}(x) \equiv t \exp i [ \gamma_{n+1}(x) - \gamma_{n}
(x) ] $. 
The random potential is thus eliminated but the hopping
matrix elements are now modulated by random phases.

The tight-binding model can be rewritten in second-quantized
language by introducing $ c^{R \dagger}_{n} $ and $ c^{R}_{n} $
which create and annihilate fermions on site $ n $ of the
tight-binding lattice and which evolve in the chiral
time direction according to the
Hamiltonian
\begin{equation}
h_{{\rm 1d}}^{R} (x) = \sum_{n} \{ t_{n}(x) c^{R \dagger}_{n}
c^{R}_{n+1} + t_{n-1}^{*} (x) c_{n}^{R \dagger} c^{R}_{n-1} 
\}.
\end{equation}
In this language the Green's function is given by 
\begin{equation}
G^{R}(n,x;n',x') = <0| c^{R}_{n} P 
\exp \left( i \int_{ x'}^{x}
d x_{1} h_{{\rm 1d}} ( x_{1} ) \right) c_{n'}^{R \dagger} 
| 0 >
\end{equation}
for $ x > x' $. The symbol $ P \exp ( ) $ denotes a 
time-ordered exponential.
Note that there is no vacuum amplitude in the denominator
because we are calculating the single particle green's 
function rather than the propagator for a single
particle added to a filled fermi sea, which is the object
usually studied in many-body physics\cite{fetter}. The absence of
a denominator in eq (4) is a crucial simplification
that allows disorder averaging as discussed below.

The complex conjugate of the Green's function can be
calculated from an expression analogous to eq (4)
by introducing conjugate fermions, $ c^{A \dagger} $
and $ c^{A} $ which evolve according to a conjugate 
hamiltonian (obtained from eq (3) by making the replacements
$ c^{R} \rightarrow c^{A} $, $ t \rightarrow - t^{*} $
and $ t^{*} \rightarrow - t $). 

Localization, or its absence, is established by
calculation of the disorder-averaged diffuson propagator $ | G^{R}_{E} 
( n, x; n', x' ) |^{2} $. For this purpose it is 
neccessary to simultaneously introduce both sets of
fermions evolving according to the total hamiltonian
\begin{equation}
h_{{\rm 1d}} ( x )  =  h^{R}_{{\rm 1d}} + h^{A}_{{\rm 1d}}
 =  \sum_{n} \{ t_{n} A_{n} + t_{n}^{*} A_{n}^{\dagger} \}.
\end{equation}
Here $ A_{n} = c_{n}^{R \dagger} c_{n+1}^{R} - c^{A \dagger}_{n+1}
c^{A}_{n} $. The diffuson is then given by
\begin{eqnarray}
| G(n,x;n',x') |^{2}  
= & & \nonumber \\
<0| c_{n}^{A} c_{n}^{R} P \exp
\left( i \int_{x'}^{x} d x_{1} h_{{\rm 1d}} (x_{1}) \right) 
c_{n'}^{R \dagger} c_{n'}^{A \dagger} | 0 >. & &
\end{eqnarray}
Eq (6) provides an exact formal expression for calculating
the diffuson for a given disorder. 
The task now is to average over different
realizations of the disorder potential which is taken
to be gaussian white noise 
with correlations
$ [ t^{*}_{n} (x) t_{m} (x') ]_{{\rm imp}} = D \delta_{n,m}
\delta( x - x' ) $ and
$ [ t_{n}(x) ]_{{\rm imp}} = [ t_{n}(x) t_{m} (x') ]_{{\rm imp}} = 0 $
\cite{balents,kim,matthew}, where $ [ \ldots ]_{{\rm imp}} $
denotes an average over disorder. Due to the absence
of a denominator in eq (6) this average is easily
performed. It is neccessary only to evaluate
\begin{equation}
\left[ P \exp \left( i \int_{x'}^{x} d x_{1} h_{{\rm 1d}} (x_{1}) 
\right) \right]_{{\rm imp}} 
= \exp \{ - \overline{h}_{{\rm int}} (x - x') \}
\end{equation}
where
\begin{equation}
\overline{h}_{{\rm int}} = \frac{1}{2} D \sum_{n}
( A_{n}^{\dagger} A_{n} + A_{n} A_{n}^{\dagger}).
\end{equation}
Eq (8) can be verified by expanding the exponentials
in eq (7). The rough content of eq (7-8) is that for
calculating averages, the fermions may be taken to
evolve according to an effective Hamiltonian 
$ \overline{h}_{{\rm int}} $ which is not random and
does not depend on $ x$. It is an interacting
hamiltonian since $ A_{n} $ is bilinear.

Consider a state in which a single site $ n $ is 
simultaneously occupied by R and A fermions$-$below
this state will be identified as a magnon localized
at site $ n $. The effect of $ \overline{h}_{{\rm int}} $
on such a state is to cause both fermions to hop
together onto a neighbouring site. This physics can
be brought out clearly by defining
$ J_{n}^{z} \equiv \frac{1}{2} ( c^{R \dagger}_{n} c^{R}_{n}
- c^{A}_{n} c^{A \dagger}_{n} ) $, $ J^{+}_{n} \equiv
J^{x}_{n} + i J^{y}_{n} \equiv c^{R \dagger}_{n} c^{A \dagger}_{n} $,
$J^{-}_{n} \equiv (J^{+}_{n})^{\dagger}$, which satisfy the
su(2) algebra, and $ N_{n} \equiv c^{R \dagger}_{n} c^{R}_{n} 
+ c^{A}_{n} c^{A \dagger}_{n} $, which commutes with all the
$ J$'s. In terms of these operators
\begin{equation}
\overline{h}_{{\rm int}}  = D \sum_{n}
\left( N_{n} - \frac{1}{2} N_{n} N_{n+1} -
2 \vec{J}_{n} \vec{J}_{n+1} \right)
\end{equation}
$-$evidently a Heisenberg ferromagnet. As usual the
vacuum $ |0> $ is the ground state and exact 
low-lying excitations are magnons obtained by
constructing plane-waves from the localized
magnons mentioned above. Explicitly, a magnon
of wave-vector $k$ is given by $ \sum_{n}
J^{+}_{n} \exp i k n |0> $ and has eigenvalue
$ 2 D ( 1 - \cos k ) $.

The exact diffuson propagator can now be straightforwardly
calculated by substituting eq (7) in eq (6) and expanding
$ c^{R \dagger}_{n'} c^{A \dagger}_{n'} |0> $ and
$ <0| c^{A}_{n} c^{R}_{n} $ in terms of magnons. The
result is 
\begin{eqnarray}
\left[ |G(n,x;n',x')|^{2} \right]_{{\rm imp}} 
= \theta( x - x' ) \times & & \nonumber \\
\int_{- \pi}^{+ \pi} \frac{ d k }{ 2 \pi } 
\exp{ i k (n - n') } 
\exp \{ - 2 D ( 1 - \cos k ) ( x - x' ) \} & & 
\end{eqnarray}
in agreement with eq (98) of Balents {\em et al.} \cite{matthew}.
The physical content of eq (10) is that the electrons move
ballistically in the chiral direction and diffuse in the
transverse direction (diffusion constant $ = D $); this is
revealed, e.g., by using eq (10) to calculate the
density response function \cite{forster}. 

Next consider the directed wave problem. The electron is
assumed to be initially localized at the origin of the co-ordinate
system $ n' = 0, x'=0 $. After it moves ballistically in the
chiral direction to a location $ x $, the amplitude to be in layer
$ n $ is given by $ G^{R}( n, x; n'=0, x'=0 ) $ and the position
of the wave-packet centre $ < n > = \sum_{n} n 
| G^{R} (n,x; 0, 0) |^{2} $. By symmetry, evidently 
$ [ <n>]_{{\rm imp}} = 0 $; and the mean-square deflection of 
the wave-packet centre is therefore
\begin{equation}
[ <n>^{2} ]_{{\rm imp}} = \sum_{n,m} n m [ |G(n,x;0,0) |^{2}
|G(m,x;0,0)|^{2} ]_{{\rm imp}}.
\end{equation}
The large $ x $ asymptotic behaviour of $[ <n>^{2} ]_{{\rm imp}} $
is desired.

To perform this calculation it is neccessary to introduce {\em two}
sets each of $ R $ and $ A $ fermions ($ c^{R}, c^{A}, d^{R}, d^{A} $)
which evolve according to the hamiltonian of eq (5) but with
$ A_{n} \rightarrow c_{n}^{R \dagger} c_{n+1}^{R} - c^{A \dagger}_{n+1}
c^{A}_{n} + ( c \rightarrow d ) $. Repeating the previous
arguments, 
\begin{eqnarray}
[ |G(n,x;0,0) |^{2}
|G(m,x;0,0)|^{2} ]_{{\rm imp}} = & & \nonumber \\
 <0| c^{A}_{n} c^{R}_{n} d^{A}_{m}
d^{R}_{m} \exp \{ - \overline{h}_{{\rm int}} x \}
d^{R \dagger}_{0} d^{A \dagger}_{0} c^{R \dagger}_{0}
c^{A \dagger}_{0} | 0 >. & &
\end{eqnarray}
Here $ \overline{h}_{{\rm int}} $ is given by eq (8) but
with $ A_{n} $ redefined as above. The important states are 
(localized) two-magnon states which are of two kinds. In the first
type a $ c^{R}-c^{A} $ pair occupies one site while a
$ d^{R}-d^{A} $ pair occupies another; in the second
type $ c^{R} $ is paired with $ d^{A} $ and $c^{A} $ with
$ d^{R} $ \cite{footnote2}. These states are closed under
the action of $ \overline{h}_{{\rm int}} $ which generally
causes paired fermions to hop together to a neighbouring
site. An exception is when the two pairs occupy adjacent
sites, in which case the fermions may switch partners
and a state of one kind is transformed into the other.
Thus the two-magnon states define a sort of two-body
problem with a contact interaction of a kind familiar
from ordinary ferromagnetism \cite{bethe}. Following
the standard method explicit forms for the two-magnon
eigenstates are obtained which can be used to straightforwardly
evaluate eq (11-12); details will be given elsewhere \cite{me}. 
Contrary to the naive assumption that the centre undergoes a random
walk, and hence $ [ <n>^{2} ]_{{\rm imp}} \sim x $,
exact calculation reveals that
$ [ <n>^{2} ]_{{\rm imp}} = \sqrt{ (2 D x )/\pi } $.

Finally consider multilayers that are of finite
size in both chiral and transverse direction thus
permitting electrons to wind around the sample in
the chiral direction. During each circumnavigation,
the electrons will typically diffuse a distance
$ \sqrt{ D C } $ in the transverse direction. Consequently
there are two distinct regimes depending on whether $ N \ll
\sqrt{ DC } $ (zero-dimensional limit) or $ N \gg \sqrt{D C} $
(one-dimensional limit). Here $ N = $ number of layers
in the multilayer. Finite sized samples are difficult
to analyse due to the complex interference produced
by winding paths. However, the conductance fluctuations
can be calculated using diagrammatic perturbation
theory within the approximation normally used for
diffusive electrons \cite{stone}. The result
is $ [ (\delta g )^{2}]_{imp} = A (e^{2}/h)^{2} $  
(one-dimensional limit) and $ [ ( \delta g )^{2} ]_{imp}
= A' \{ ( C D )/ N^{2} \} (e^{2} / h)^{2} $ (zero-dimensional
limit). Here $ A $ and $ A'$ are constants of order
unity which we have not computed explicitly
\cite{footnote3}. Conductance fluctuations are
of order $ e^{2}/h $ in the one-dimensional 
limit as they are for ordinary metallic grains. The larger
non-universal result for the zero-dimensional
limit may be interpreted as follows: The electron
moves a distance $ N^{2}/D \ll C $ in the chiral direction
before it diffuses into the phase-randomizing 
probes. The sample therefore breaks up into
$ C D/N^{2} $ incoherent blocks, each with
independent conductance fluctuations of order
$ e^{2}/h $. Although the perturbative
results given here have an appealing physical
interpretation, the validity of these results
deserves further study via simulation or
non-perturbative analysis (which may be possible
in the zero-dimensional limit \cite{igor2}).
By the familiar ``ergodic hypothesis''
of mesoscopic physics \cite{stone} we expect the
statistical fluctuations of the conductance
calculated here would be experimentally manifested
as fluctuations in the conductance of a given
specimen when a tunable parameter, e.g., the magnetic
field, is varied. 

It is interesting to compare the method of this
paper with other field theory representations
of the chiral model. Because single particle
properties are being calculated, eq (3$-$8) would
remain valid even if the fermions were replaced
by bosons ($ c^{R}, c^{A} \rightarrow b^{R}, b^{A} $)
yielding an interacting boson representation of the chiral
model rather than the Heisenberg ferromagnet analysed
here. If fermions {\em and} a redundant set of bosons
are introduced, the supersymmetric representation of
Balents {\em et al.} \cite{matthew} results. It is very convenient 
to be rid of the unneeded bosons, e.g., while analysing
the directed wave problem; but it is important to 
emphasize that the bosons are optional only for
the infinite system. For a finite circumference
the boundary conditions on eq (1) change and it
becomes neccessary to introduce bosons whether
operator methods (this paper) or functional methods
(Balents {\em et al.}) are used. 

In summary, the main results of this paper are:
a mapping of chiral waves onto a Heisenberg 
spin-chain with ferromagnetic coupling; 
calculation of the diffuson 
propagator, eq (10), which shows the surface
of a quantum hall multi-layer to be an anisotropic
metal in the infinite size limit; 
diagrammatic calculation of conductance
fluctuations in two diffusive mesoscopic regimes 
for finite sized multilayers;
and the exact result that the mean-square deflection
of a directed wave grows as the square root of the
distance it has propagated.

It is a pleasure to thank Leon Balents and Matthew Fisher
for illuminating discussions and for patient
explanation of the contents of ref
\cite{balents,matthew} and Yi Kuo Yu for 
discussion of conductance fluctuations.  
Thanks are also due to Matthew
Fisher for hospitality at the ITP Santa Barbara where
this work was initiated.  This work was supported by
NSF Grant PHY 94-07194, startup funds from CWRU
and an Alfred P Sloan Research fellowship.

\vspace{50mm}

\figure{Fig 1. A quantum hall multilayer.}

\end{multicols}


\begin{references}
\bibitem{footnote2} The states for which a single site is
occupied by all four fermions are special and may be considered
to be of either type.
\bibitem{footnote3} To be precise, we have explicitly
evaluated the diagram in fig 5(a) of ref \cite{stone}
which yields $ A = 4 \zeta(4)/\pi^{4}$ and $ A' = 
2 \zeta(2)/ \pi^{2} $. Diagrams 5 (b-e)
should yield the same parametric dependence
but the coefficients have not been computed.
\end{references}
\end{document}